\documentclass[12pt,notitlepage,onecolumn,showpacs]{revtex4-1}
\usepackage{amsmath}
\usepackage{amsfonts}
\usepackage{amssymb}
\usepackage{times,fullpage}
\usepackage{comment,bm}
\usepackage{array,graphicx}

\graphicspath{{./figures/}}

\begin{document}

\newcommand{\nwc}{\newcommand}
\nwc{\vs}{\vspace}
\nwc{\hs}{\hspace}
\nwc{\la}{\langle}
\nwc{\ra}{\rangle}
\nwc{\nn}{\nonumber}
\nwc{\Ra}{\Rightarrow}
\nwc{\wt}{\widetilde}
\nwc{\lw}{\linewidth}
\nwc{\ft}{\frametitle}
\nwc{\ben}{\begin{enumerate}}
\nwc{\een}{\end{enumerate}}
\nwc{\bit}{\begin{itemize}}
\nwc{\eit}{\end{itemize}}
\nwc{\dg}{\dagger}
\nwc{\mA}{\mathcal A}
\nwc{\mD}{\mathcal D}
\nwc{\mB}{\mathcal B}

\nwc{\Tr}[1]{\underset{#1}{\mbox{Tr}}~}
\nwc{\pd}[2]{\frac{\partial #1}{\partial #2}}
\nwc{\ppd}[2]{\frac{\partial^2 #1}{\partial #2^2}}
\nwc{\fd}[2]{\frac{\delta #1}{\delta #2}}
\nwc{\pr}[2]{K(i_{#1},\alpha_{#1}|i_{#2},\alpha_{#2})}
\nwc{\av}[1]{\left< #1\right>}

\nwc{\zprl}[3]{Phys. Rev. Lett. ~{\bf #1},~#2~(#3)}
\nwc{\zpre}[3]{Phys. Rev. E ~{\bf #1},~#2~(#3)}
\nwc{\zpra}[3]{Phys. Rev. A ~{\bf #1},~#2~(#3)}
\nwc{\zjsm}[3]{J. Stat. Mech. ~{\bf #1},~#2~(#3)}
\nwc{\zepjb}[3]{Eur. Phys. J. B ~{\bf #1},~#2~(#3)}
\nwc{\zrmp}[3]{Rev. Mod. Phys. ~{\bf #1},~#2~(#3)}
\nwc{\zepl}[3]{Europhys. Lett. ~{\bf #1},~#2~(#3)}
\nwc{\zjsp}[3]{J. Stat. Phys. ~{\bf #1},~#2~(#3)}
\nwc{\zptps}[3]{Prog. Theor. Phys. Suppl. ~{\bf #1},~#2~(#3)}
\nwc{\zpt}[3]{Physics Today ~{\bf #1},~#2~(#3)}
\nwc{\zap}[3]{Adv. Phys. ~{\bf #1},~#2~(#3)}
\nwc{\zjpcm}[3]{J. Phys. Condens. Matter ~{\bf #1},~#2~(#3)}
\nwc{\zjpa}[3]{J. Phys. A ~{\bf #1},~#2~(#3)}
\nwc{\zpjp}[3]{Pramana J. Phys. ~{\bf #1},~#2~(#3)}

\title{Fluctuation Theorems of work and entropy in Hamiltonian systems}
%\author{Sourabh Lahiri and A. M. Jayannavar}
%\date{}
 \author{Sourabh Lahiri$^1$} 
 \email{sourabhlahiri@gmail.com}
 \author{A. M. Jayannavar$^{2,3}$}
 \email{jayan@iopb.res.in}
 \affiliation{$^1$International Centre for Theoretical Sciences, TIFR, Survey no. 151, Sivakote Village, Hesaraghatta Hobli, Bengaluru 560089, India\\
   $^2$Institute of Physics, Sachivalaya Marg, Bhubaneswar 751005, India\\
   $^3$Homi Bhabha National Institute, Training School Complex, Anushakti Nagar, Mumbai 400085, India}

 \begin{abstract}
 The Fluctuation Theorems are a group of exact relations that remain valid irrespective of how far the system has been driven away from equilibrium. Other than having practical applications, like determination of equilibrium free energy change from nonequilibrium processes, they help in our understanding of the Second Law and the emergence of irreversibility from time-reversible equations of motion at microscopic level. A vast number of such theorems have been proposed in literature, ranging from Hamiltonian to stochastic systems, from systems in steady state to those in transient regime, and for both open and closed quantum systems. In this article, we discuss about a few such relations, when the system evolves under Hamiltonian dynamics.
 \end{abstract}
% \pacs{}

\maketitle
\section{Introduction}

The fluctuation theorems (FTs) are a group of exact relations that remain valid even when the system of interest is driven far away from equilibrium \cite{eva93_prl,eva94_pre,jar97_prl,cro98_jsp,cro99_pre,sei12_rpp}. For driven systems fluctuations in heat, work and entropy are not mere background noise, but satisfy strong constraints on the probability distributions of these fluctuating quantities. These relations are of fundamental importance in non-equilibrium statistical mechanics. Intensive research has been done in this direction in order to find such relations for thermodynamic quantities like work, heat or entropy changes. They have resulted in conceptual understanding of how irreversibility emerges from reversible dynamics and of the second law of Thermodynamics \cite{jar10_arcmp,gom08_epl}. These theorems lead to the fact that the Second Law holds for average quantities. However, there are atypical transient trajectories in phase space which violate second law . Two fundamental ingredients play a decisive role in the foundations of FT - the principle of micro reversibility and the fact that thermal equilibrium is described by Gibbs canonical ensemble. Moreover, some of these relations have been found useful for practical applications like determining the change in equilibrium free energy in an irreversible process. Numerous FTs have been put forward in the last two decades \cite{sei12_rpp}. Some of them are valid when the system is in a non equilibrium steady state, while the others are valid in the transient regime. Fluctuation theorems have been proposed for Hamiltonian as well as stochastic dynamics, and for quantum systems(both closed and open ones). Some of them have been tested experimentally.

To understand the significance of the FTs, we first need to appreciate the fact that the phase space trajectories of small systems (at meso or nano scales) are stochastic trajectories, because thermal fluctuations play a dominant role in their dynamics. In particular, if the evolution is Hamiltonian, then the stochasticity comes from the fact that the initial state of the system is sampled from some distribution that is not simply at delta-function. As a result, thermodynamic quantities like work, heat or entropy change are also stochastic and follow distributions instead of having a single value. The FTs provide stringent conditions on the symmetries of these distributions. Furthermore, the inequalities encountered in the Second Law can be readily obtained as corollaries from the FT relations. Thus, these theorems are stronger relations than the Second Law. We now move on to describe the FTs for work.

\section{Fluctuation theorems for work}

These theorems include the Crooks work FT \cite{cro98_jsp,cro99_pre} and the Jarzynski Equality \cite{jar97_prl,jar97_pre}. They show that the equilibrium free energy difference between the final and initial values of the external parameter can be computed by measuring the nonequilibrium work done, instead of having to perform the experiment quasistatically. This has a lot of practical significance.

Let $z\equiv (\bm q,\bm p)$  denote the position and momenta of the system degrees of freedom. Initially, the system is prepared at equilibrium with a heat bath at temperature $T$, and at time $t=0$ the bath is disconnected and an external parameter $\lambda(t)$ that acts on the system is switched on. Let $H_S(z,t)$ be the system Hamiltonian that depends explicitly on time, so that the initial state distribution of the system is given by
\begin{align}
  \rho_{\lambda_0}(z_0) = \frac{e^{-\beta H_S(z_0)}}{Z_0},
\end{align}
$Z_0$ being the initial partition function defined by
\begin{align}
  Z = \int dz_0 e^{-\beta H_S(z_0)}.
\end{align}
  At the end of the process, at time $t=\tau$, the parameter is turned off.
The probability of a trajectory $z(t)$ in phase space is given by $P[z(t)]$.

\begin{figure}[!h]
  \centering
  \includegraphics[width=8cm]{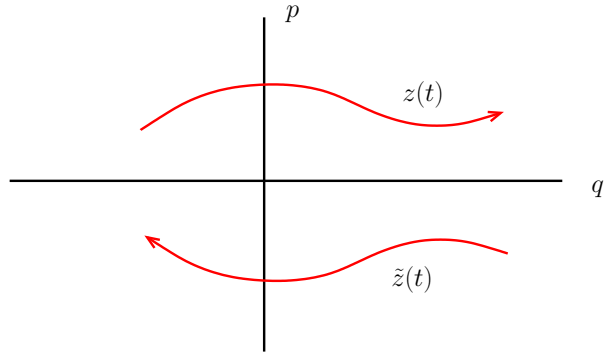}
  \caption{A typical forward and its correponding reverse trajectory}
  \label{fig:reverse_traj}
\end{figure}

Correspondingly, a time-reversed process is defined which is described by the parameter $\lambda(\tau-t)$ acting on the system that is initially at equilibrium with the final value of the forward protocol. The reverse trajectory corresponding to the forward trajectory $z(t)$ is defined as $\tilde z(t) = z^*(\tau-t)$. Here, the asterisk implies that all the velocity variables switch signs. A typical forward and its corresponding reverse trajectory are depicted in figure \ref{fig:reverse_traj}.
The initial state distribution of the system for the reverse process is then given by
\begin{align}
  \tilde\rho_{\lambda_\tau}(z_\tau) = \frac{e^{-\beta H_S(z^*_\tau)}}{Z_\tau}
\end{align}
 The probability of a reverse trajectory in the reverse process is given by $\tilde P[\tilde z(t)]$.
If we take the ratio between $P[z(t)]$ and $\tilde P[\tilde z(t)]$, then we get
\begin{align}
  \frac{P[z(t)]}{\tilde P[\tilde z(t)]} = \frac{\rho_{\lambda_0}(z_0)}{\tilde\rho_{\lambda_\tau}(z_\tau)} = e^{\beta[H_S(z_\tau)-H_S(z_0)-\Delta F]},
\end{align}
where we have used the fact that there is a one-to-one correspondence between the entire trajectory and its initial point in a deterministic evolution. We have further assumed that the equilibrium Hamiltonian is symmetric under time reversal: $H_S(z^*_\tau) = H_S(z_\tau)$. $\Delta F$ is the free energy change of the system: $\Delta F = k_BT\ln\frac{Z_0}{Z_\tau}$.  Since the total work $W$ done on the system must be equal to the change in the Hamiltonian: $W = H_S(z_\tau)-H_S(z_0)$, we can write
\begin{align}
  \frac{P[z(t)]}{\tilde P[\tilde z(t)]} = e^{\beta(W-\Delta F)}.
  \label{microrev}
\end{align}
This is the condition of microscopic reversibility. We remark that the same relation holds even for stochastic dynamics, although the derivation of the condition of microscopic reversibility is not so simple in that case.

Now we will prove two FTs: one is called the Crooks FT, and the other is known as Jarzynski equality. The Crooks FT directly follows from Eq. \eqref{microrev}:
\begin{align}
  P(W) &= \int \mD[z(t)]\delta(W-W[z(t)])P[z(t)] \nn\\
       &= \int \mD[z(t)]\delta(W-W[z(t)]) \tilde P[\tilde z(t)] e^{\beta(W-\Delta F)} \nn\\
       &= e^{\beta(W-\Delta F)} \int \mD[z(t)]\delta(W+W[\tilde z(t)]) \tilde P[\tilde z(t)]\nn\\
       &= e^{\beta(W-\Delta F)}\tilde P(-W).
         \label{CFT}
\end{align}
This is the Crooks FT for work.
In the third line, we have used the fact that $W[z(t)]=-W[\tilde z(t)]$. Now, if we multiply both sides of the above relation by $e^{-\beta(W-\Delta F)}$ and integrate over $W$, then, using the normalization condition for $\tilde P(-W)$, we obtain
\begin{align}
  \av{e^{-\beta W}} = e^{-\beta\Delta F}.
  \label{JE}
\end{align}
Thus, from \eqref{CFT} and \eqref{JE}, we find that the change in the equilibrium free energy of the system can be computed from nonequilibrium work measurements, which removes the necessity of using a quasi-static process for obtaining $\Delta F$. This has a lot of practical importance.

We will now show that the JE leads to the Maximum Work Theorem, an alternative statement for the Second Law. First, we prove the relation $\av{e^r}\ge e^{\av r}$ for some variable $r$, using the identity $e^r\ge 1+r$ for all values of $r$, the equality being satisfied only at $r=0$. The proof goes as follows \cite{cha}:
\begin{align}
  \av{e^r} &= \av{e^{r+\av r-\av r}} = e^{\av r}\av{e^{r-\av r}}\nn\\
           &\ge e^{\av r}\av{1+r-\av r} = e^{\av r}.
\end{align}
This is the so-called Jensen's inequality for exponential functions. Using this in \eqref{JE}, we find that
\begin{align}
  e^{-\beta\Delta F} \ge e^{-\beta\av W} ~~~\Ra~~~ \av W \ge \Delta F.
\end{align}
Multiplying both sides by $-1$, we have $-\av{W}\le -\Delta F$, which is the statement of the maximum work theorem:
\emph{For all thermodynamic processes between some initial and final states, the delivery of work is a maximum for a reversible process.} This maximum work extracted is given by the free energy difference $-\Delta F$.

\section{Experimental tests of the work fluctuation theorems}

The CFT as well as JE have been tested experimentally on various systems: RNA hairpin \cite{rit05_nat,rit05_pt}, dragged colloidal particle \cite{eva02_prl}, torsion pendulum \cite{cil10_jsm}, electric circuits \cite{cil04_prl,cil05_pre}, etc.  Let us describe one of them, namely the experiments performed on RNA hairpin, in order to provide a visualization of the applications of these theorems.
Note, however, that the experiments deal mostly with stochastic dynamics, since the system is not isolated from the reservoir. However, as mentioned before, the FTs have been derived for stochastic dynamics as well. In this sense, these theorems are very general.

The experimental setup is as follows: the two ends of an RNA hairpin are connected to two beads, one of whose position has been fixed by using a strong laser trap. The other bead is moved by an actuator, according to a protocol $\lambda(t)$, which for simplicity was chosen to be a linear function of time. Initially the RNA hairpin is at equilibrium in the folded state. The experiment was repeated a large number of times, and $\Delta F$ was computed using Eq. \eqref{JE}. The computed value agreed very well with the independent estimate using a quasi-static process.

Verification of CFT was carried out by generating ensemble of realizations of both the forward and the reverse process. In the present context, the reverse process consists of reducing the distance between the two ends of the RNA molecule, so that it is a \emph{refolding} process, as opposed to the \emph{unfolding} or forward process. Of course, the initial distribution of the reverse process has to be the thermal one, corresponding to the stretched RNA hairpin with the end-to-end distance being the same as the one reached at the end of the forward process. Distributions of work were constructed in the forward and the reverse processes, and $P(W)$ and $\tilde P(-W)$ were plotted together. As per Eq. \eqref{CFT}, when $P(W)=\tilde P(-W)$, then $W=\Delta F$. Thus, the intersection of the two distributions should provide the value of the change in free energy. Such distributions were constructed for different speeds of the driving protocol, keeping the initial and final end-to-end distance of the RNA hairpin the same. It was observed that for all pairs of distributions $P(W)$ and $\tilde P(-W)$, the intersection points were placed on the same vertical line, which is consistent with the fact that the free energy difference must be the same in all the cases, since the initial and final values of the protocols were same.

\section{Fluctuation theorems for change in total entropy}

We now discuss the FTs for the entropy change of an isolated system. These theorems come in two forms: the detailed FT which is valid only for transitions between steady states, and an integral FT that is valid for all systems and leads to the well-known statement of the thermodynamic Second Law \cite{sei05_prl,sei08_epjb}.

Let us consider a system that is prepared with an initial distribution $p(z_0,0)$ which need not be a thermal distribution. It is then evolved under a protocol $\lambda(t)$ up to time $t=\tau$, and its final distribution is given by $p(z_\tau,\tau)$. The system's entropy at any point in time is defined as
\begin{align}
  s(t) = -\ln p(z_t,t),
\end{align}
so that over an ensemble the average entropy is given by the Gibbs form:
\begin{align}
  \av{s(t)} = -\int dz_t ~p(z_t,t)\ln p(z_t,t).
  \label{entropy}
\end{align}
The total change in entropy in the process is then given by the difference between the system's entropy at the final and the initial times:
\begin{align}
  \Delta s(t) = \ln \frac{p(z_0,0)}{p(z_\tau,\tau)}.
\end{align}
Now, the probability of $\Delta s(t)$ is given by (see \cite{lah13_pramana} for the corresponding derivation for isolated quantum systems)
\begin{align}
  P(\Delta s) &= \int dz_0dz_\tau~\delta\bigg[\Delta s - \ln \frac{p(z_0,0)}{p(z_\tau,\tau)}\bigg]P[z_0,z_\tau] \nn\\
              &= \int dz_0dz_\tau~\delta\bigg[\Delta s - \ln \frac{p(z_0,0)}{p(z_\tau,\tau)}\bigg]P(z_\tau|z_0)p(z_0,0)\nn\\
              &= \int dz_0dz_\tau~\delta\bigg[\Delta s - \ln \frac{p(z_0,0)}{p(z_\tau,\tau)}\bigg]P(z_\tau|z_0)p(z_\tau,\tau)\frac{p(z_0,0)}{p(z_\tau,\tau)}\nn\\
              &= e^{\Delta s} \int dz_0dz_\tau~\delta\bigg[\Delta s - \ln \frac{p(z_0,0)}{p(z_\tau,\tau)}\bigg] P(\tilde z_0|\tilde z_\tau)\tilde p(\tilde z_\tau,\tau)\nn\\
              &= e^{\Delta s} \int dz_0dz_\tau~\delta\bigg[\Delta s + \ln \frac{\tilde p(\tilde z_\tau,\tau)}{\tilde p(\tilde z_0,0)}\bigg] P(\tilde z_0|\tilde z_\tau)\tilde p(\tilde z_\tau,\tau)\nn\\
              &= \tilde P(-\Delta s)e^{\Delta s}.
\end{align}
where we have used the fact that under deterministic evolution, $P(z_\tau|z_0) = P(\tilde z_0|\tilde z_\tau)$, and we have assumed that the initial and the final distributions are symmetric under time-reversal: $p(z_0,0) = \tilde p(\tilde z_0,0)$, and $p(z_\tau,\tau) = \tilde p(\tilde z_\tau,\tau)$. This is the detailed fluctuation theorem for entropy change, and the corresponding integral FT is obtained by simply multiplying both sides by $e^{-\Delta s}$ and integrating over $\Delta s$:
\begin{align}
  \av{e^{-\Delta s}}=1.
  \label{IFT}
\end{align}
The RHS is unity due to normalization of $\tilde P(-\Delta s)$. Once again, just as in the case of JE, we can apply the Jensen's inequality to the above theorem, which would give the statement of Second Law: $\av{\Delta s}\ge 0$, i.e. \emph{the average entropy of an isolated system never decreases with time}.

{\bf Remark:} With the definition \eqref{entropy}, it can be shown that the entropy $s(t)$ is a constant, independent of time \cite{schwabl}: $\av{\Delta s}=0$, so that the inequality in the Second Law statement becomes inconsequential. However, the change in entropy along individual realizations can be non-zero, which is why \eqref{IFT} is a non-trivial result. Further, the same relation is true even for stochastic system, where the system itself and the heat reservoir together form an isolated system. Defining the system entropy and bath entropy changes separately and adding them up gives the total change in entropy of the full isolated system. In this case, the second law statement becomes $\av{\Delta s_{sys}}+\av{\Delta s_{bath}}\ge 0$, where the average total entropy will increase with time, unless the process is an equilibrium one. Note that this way of defining total entropy is different from simply taking the negative logarithm of the state distribution of the combined system (consisting of the system and the heat bath).

\section{Work fluctuation theorems in quantum systems}

The work fluctuation theorems have been proven for both open and closed quantum systems \cite{han07_jpa,han09_jsm,han10_prl,han11_rmp,lah12_pramana,lah13_pramana}. The fact that work can only be defined (see below) by performing two-point measurements on the Hamiltonian of the system was clarified in \cite{han07_pre}. Here we provide the derivation for a system which is initially at equilibrium with a heat bath, but at time $t=0$ it has been disconnected from the bath and allowed to evolve unitarily under the action of an external time-dependent protocol $\lambda(t)$, just as in the case of a classical isolated system.

Let us consider an isolated quantum system undergoing unitary evolution under a time-dependent Hamiltonian $\hat H(t)$. The initial state of the system has been sampled from canonical distribution and thereafter the system and heat bath have been disconnected. The initial density operator is thus given by
\begin{align}
  \hat\rho(\lambda_0) = \frac{e^{-\beta \hat H(0)}}{Z_0},
\end{align}
where $Z=\mbox{Tr}~e^{-\beta \hat H}$. At time $t=0$, a projective measurement is performed on the Hamiltonian. Let the state of the system collapse to the eigenstate $|n\ra$ with an eigenvalue $E_n$. Then the system is allowed to evolve unitarily till time $t=\tau$ when another projective measurement is performed. Let the final eigenstate be $|m\ra$ and the corresponding eigenvalue be $E_m$. The work done on the system is simply equal to the change in the energy of the system:
\begin{align}
  W = E_m-E_n.
\end{align}
The density operator at the beginning of the reverse process is given by
\[
  \hat\rho(\lambda_\tau) = \frac{e^{-\beta \hat H(\tau)}}{Z_\tau}.
\]
The probability of obtaining the eigenvalue $E_n$ at time $t=0$ is given by $p^{eq}_0(n) = \Tr{}[\hat\rho(\lambda_0)|n\ra\la n|] = e^{-\beta E_n}/Z$. Similarly, the probability of obtaining eigenvalue $E_m$ at the beginning of the reverse process is given by $p^{eq}_\tau(m) = \Tr{}[\hat\rho(\lambda_\tau)|m\ra\la m|] = e^{-\beta E_m}/Z_\tau$. 
The probability of work is then given by
\begin{align}
  P(W) &= \sum_{mn}\delta[W-(E_m-E_n)] P(m|n)p^{eq}_0(n)\nn\\
       &=\sum_{mn}\delta[W-(E_m-E_n)]|\la m|U_\lambda(\tau,0)|n\ra|^2 ~\frac{e^{-\beta E_n}}{Z_0}\nn\\
       &= \sum_{mn}\delta[W-(E_m-E_n)]|\la n|U_{\tilde\lambda}(0,\tau)|m\ra|^2 ~\frac{e^{-\beta (E_m-W)}}{Z_\tau}\frac{Z_\tau}{Z_0}\nn\\
       &= e^{\beta(W-\Delta F)}\sum_{mn}\delta[W-(E_m-E_n)]|\la n|U_{\tilde\lambda}(0,\tau)|m\ra|^2~\frac{e^{-\beta E_m}}{Z_\tau}\nn\\
       &= e^{\beta(W-\Delta F)}\sum_{mn}\delta[-W-(E_n-E_m)]\tilde P(n|m)p^{eq}_\tau(m)\nn\\
       &= e^{\beta(W-\Delta F)}\tilde P(-W).
\end{align}
While going from second to the third line, we have used the fact that unitary evolution is reversible in time: $|\la m|U_\lambda(\tau,0)|n\ra|^2 = |\la n|U_{\tilde\lambda}(0,\tau)|m\ra|^2$, if the external protocol is time-reversed. This is the Crooks Fluctuation Theorem for a quantum system. The Jarzynski equality follows from the CFT readily by cross-multiplication followed by integration over $W$.

\section{Extended Fluctuation Theorems in presence of feedback}

Now let us consider a system subjected to a feedback-dependent protocol. This means that the functional form of the protocol with time will depend on outcomes of measurements made on some property of the system. Such systems have been studied in \cite{sag10_prl,hor10_pre,pon10_pre,lah12_jpa,sag12_pre}. To keep our discussions simple, we will always assume that the measurements are made on the particle state in phase space for a classical system and projective measurements are made on the Hamiltonian for a quantum system. Let us first study the classical system undergoing Hamiltonian evolution. At time $t=0$, a given protocol $\lambda_0(t)$ is switched on. The system evolves under this protocol till time $t=t_m$ when a measurement is made on the state of the particle, and let the outcome be $m$ which can be different from the true state $z_m$ due to measurement inaccuracies. This outcome is obtained with probability $p(m|z_m)$. Depending on the outcome, the protocol is changed to $\lambda_{m}(t)$ which acts on the system till the final time $t=\tau$. The probability of a forward trajectory is given by
\begin{align}
  \label{Pf_feedback}
  P[z(t),m] &= p_0^{eq}(z_0)P_{\lambda_0}[z_0\to z_m]p(m|z_m)P_{\lambda_m}[z_m\to z_\tau]\nn\\
            &= P_m[z(t)]p(m|z_m).
\end{align}
The reverse process is generated by choosing one of the forward processes (corresponding to one of the values of $m$) and blindly time-reversing this protocol without applying any feedback. The probability of a reverse trajectory is
\begin{align}
  \label{Pr_feedback}
  \tilde P[\tilde z(t),m] &= \tilde P_m[\tilde z(t)]p(m).
\end{align}
Dividing \eqref{Pf_feedback} by \eqref{Pr_feedback}, we get
\begin{align}
  \label{MR_feedback}
  \frac{P[z(t),m]}{\tilde P[\tilde z(t),m]} &= e^{\beta(W[z(t),m]-\Delta F(m))+I}.
\end{align}
Here, we have used the condition of microscopic reversibility for any given set of forward and reverse protocols:
\begin{align}
  \frac{P_m[z(t)]}{\tilde P_m[\tilde z(t)]} = e^{\beta(W[z(t),m]-\Delta F(m))},
\end{align}
and have defined the \emph{mutual information} $I$ between $m$ and $z_m$ as
\begin{align}
  I = \ln \frac{p(m|z_m)}{p(m)}.
\end{align}
Thus, with \eqref{MR_feedback}, we obtain
\begin{align}
  \av{e^{-\beta(W-\Delta F)-I}} = 1.
\end{align}
Application of Jensen's inequality then leads to
\begin{align}
  \av{W-\Delta F} \ge -k_BT\av{I},
\end{align}
where
\begin{align}
  \av{I} = \int dm \int dz_m ~p(m,z_m)\ln \frac{p(m|z_m)}{p(m)}.
\end{align}
Since $\av{I}$ is a Kullback-Leibler divergence, it is always non-negative \cite{cov}, which means that the dissipated work $W_d\equiv W-\Delta F$ can become negative on average. In other words, on an average, more work can be extracted from the system than the change in free energy. This is the modified Second Law in presence of information gain and feedback.

\section{Guessing the direction of time's arrow}

When we observe a movie of some process, we can easily make out whether the movie is run forwards or backwards. For instance, if the movie shows that broken shards of a flower vase arrange themselves into the original flower vase, then we can say with confidence that the movie is being run backward. As another example, if we find that a die that is well mixed with water is getting concentrated to form the original solid dye, then we can be sure that the movie is being run backwards. However, such a directionality in the ``time's arrow'' gets blurred when we go to mesoscopic or smaller systems, where thermal fluctuations can give rise to unexpected events. In such cases, $W>\Delta F$ does not necessarily mean that the process is running forward in time, and vice versa. In the following, we explain the quantification of the ability to guess time's arrow from a hypothetical guessing game, following the treatment in \cite{jar10_arcmp,jar08_epjb}.

The game consists of watching a movie and trying to guess whether it is being run forward or backward. Given a trajectory $z(t)$ in phase space, the likelihood that it is the forward process is given by $P[F|z(t)]$, and that of it being the reverse process run backwards is $P[R|z(t)]$. The total likelihood of the movie being run either forward or backward is unity:
\begin{align}
  \label{norm}
  P[F|z(t)] + P[R|z(t)] = 1.
\end{align}
Bayes' Theorem tells us
\begin{align}
  \label{bayesF}
  P[F|z(t)] = \frac{P[z(t)|F]P_F}{P[z(t)]}.
\end{align}
Similarly,
\begin{align}
  \label{bayesR}
  P[R|z(t)] = \frac{P[z(t)|R]P_R}{P[z(t)]}.
\end{align}
Here, $P[z(t)|F]$ is the probability of the trajectory $z(t)$ in the forward process, which is simply $P[z(t)]$ in our earlier notation. Similarly, $P[z(t)|R]$ is the probability that the trajectory $z(t)$ is observed, given that the movie of the reverse process is being run backwards, so that $P[z(t)|R] = \tilde P[\tilde z(t)]$.
$P_F$ is the prior probability of a forward process and $P_R$ is that for the reverse process. If the forward and reverse processes are shown based on the outcome of a tossed fair coin, then $P_F=P_R=1/2$. Dividing \eqref{bayesF} by \eqref{bayesR}, we obtain
\begin{align}
  \label{guess}
  \frac{P[F|z(t)]}{P[R|z(t)]} = e^{\beta(W-\Delta F)},
\end{align}
where we have used the relation for microscopic reversibility \eqref{microrev}. Now, from \eqref{norm} and \eqref{guess}, we finally arrive at the following form of the likelihood:
\begin{align}
  P[F|z(t)] = \frac{1}{1+e^{-\beta(W-\Delta F)}}.
\end{align}
Thus, if $W\gg\Delta F$, then $P[F|z(t)]\approx 1$, and we can say with a high degree of confidence that the movie is running forward in time. In the other extreme, if $W\ll\Delta F$, then $P[F|z(t)]\approx 0$, and we can say with a high degree of confidence that the actual process was the reverse one, whose movie is being run backwards.

\section{Conclusions}

In this short article, we have discussed the Fluctuation Theorems for work and entropy change for isolated systems undergoing Hamiltonian evolution in classical case or unitary evolution in the quantum case. We have seen that the concept of microscopic reversibility directly gives rise to the Crooks FT which in turn leads to the Jarzynski Equality. Use of Jensen's inequality on the JE or on the integral FT for entropy change leads to alternative statements of the Second Law. These theorems get modified if the external protocol is feedback-driven, and so does the Second Law inequality. In this situation, the extracted work can become more than the free energy difference between the final and the initial states. We have also discussed the likelihood with which one can guess correctly the direction of time's arrow by looking at the movie of a process.

\section{Acknowledgements}

One of us (AMJ) thanks DST, India for financial support (through J. C. Bose National Fellowship).


%\bibliographystyle{apsrev4-1}
%\bibliography{/home/lahiri/ref}

\end{document}